\documentclass[draftcls,3p,times]{article}
\usepackage{graphicx, subfigure}
\usepackage{amssymb}
\usepackage{amsthm}
\usepackage{dcolumn}
\usepackage{bm}
\usepackage{amsfonts,amssymb}
\newtheorem{theorem}{Theorem}[section]

\begin{document}
\pagestyle{plain}
\title
{\bf A simple yet complex one-parameter family of generalized
lorenz-like systems\thanks{This work was supported by the
France/HongKong Joint Research Scheme under grant CityU
9052002/2011-12.}}
\author{Xiong Wang$^1$\thanks{Email:wangxiong8686@gmail.com},  Juan Chen$^{1,2}$, Jun-An Lu$^2$ and Guanrong Chen$^{1}$\\
$^{1}$ Department of Electronic Engineering,\\ City University of Hong Kong, Kowloon, Hong Kong\\
$^{2}$ School of Mathematics and Statistics,\\ Wuhan University, Wuhan 430072, China \\
}

 \maketitle

%
%
%
%
\begin{abstract}
This paper reports the finding of a simple one-parameter family of
three-dimensional quadratic autonomous chaotic systems. By tuning
the only parameter, this system can continuously generate a
variety of cascading Lorenz-like attractors, which appears to be
richer than the unified chaotic system that contains the Lorenz
and the Chen systems as its two extremes. Although this new family
of chaotic systems has very rich and complex dynamics, it has a
very simple algebraic structure with only two quadratic terms
(same as the Lorenz and the Chen systems) and all nonzero
coefficients in the linear part being $-1$ except one $-0.1$
(thus, simpler than the Lorenz and Chen systems). Surprisingly,
although this new system belongs to the family of Lorenz-type
systems in some existing classifications such as the generalized
Lorenz canonical form, it can generate not only Lorenz-like
attractors but also Chen-like attractors. This suggests that
there may exist some other unknown yet more essential algebraic
characteristics for classifying general three-dimensional
quadratic autonomous chaotic systems.
\end{abstract}
\section{Introduction}

The scientific story of chaos dates back to earlier 1960s
\cite{Alorenz}, when Lorenz studied the atmospheric convection
phenomenon he found a chaotic attractor in a three-dimensional (3D)
quadratic autonomous system \cite{Alorenz2}. The now-classic Lorenz
system is described by
\begin{equation}
 \left\{
 \begin{array}{l}
 \dot{x}=\sigma \left( y-x\right) \\
 \dot{y}=rx-y-xz \\
 \dot{z}=-bz+xy,%
 \end{array}%
 \right.
\end{equation}
which is chaotic when $\sigma =10,\ r=28,\ b=\frac{8}{3}.$

Ever since then, many researchers were wondering and pondering
whether the discovery of the Lorenz system was just a lucky
incident or there actually exist other closely related systems
around it?

In 1999, from an engineering feedback anti-control approach, Chen
provided a certain answer to this question by his finding of a new
system \cite{Achen1999,Aueta2000}, lately referred to as the Chen
system by others, described by
\begin{equation}
 \left\{
 \begin{array}{l}
 \dot{x}=a(y-x)\\
 \dot{y}=(c-a)x-xz+cy \\
 \dot{z}=-bz+xy,%
 \end{array}%
 \right.
\end{equation}
which is chaotic when $a=35,\ b=3,\ c=28$. Whereafter, it has been
proved that Chen's attractor exists \cite{AChenexists} and the Chen
system displays even more sophisticated dynamical behaviors than the
Lorenz system \cite{Achendynamical2003}.

It is also interesting to recall a unified chaotic system, which was
constructed to encompass both the Lorenz system and the Chen system
\cite{Alu2002b}. This unified chaotic system is by nature a convex
combination of the two systems, and is described by
\begin{equation}\label{unified}
 \left\{
 \begin{array}{l}
 \dot{x}=(25\alpha+10)(y-x)\\
 \dot{y}=(28-35\alpha)x-xz+(29\alpha-1)y \\
 \dot{z}=-\frac{\alpha+8}{3}z+xy.
 \end{array}
 \right.
\end{equation}
When $\alpha=0$ it is the Lorenz system while with $\alpha=1$ it
is the Chen system, and moreover for any $\alpha\in(0,1)$ the
system remains to be chaotic.

As a result of several years of continued research endeavor along
the same line, a family of generalized Lorenz systems was found and
characterized \cite{CFCV1994,CFCV2002a,CFCV2002b,CFCV2005}, which is
defined through the so-called generalized Lorenz canonical form, as
follows:
\begin{eqnarray}
 \left(
 \begin{array}{c}
 \dot x \\
 \dot y \\
 \dot z \\
 \end{array}
 \right)= \left(
 \begin{array}{ccc}
 a_{11} & a_{12} & 0 \\
 a_{21} & a_{22} & 0 \\
 0 & 0 & a_{33} \\
 \end{array}
 \right)\left(
 \begin{array}{c}
 x \\
 y \\
 z \\
 \end{array}
 \right)+
 \left(
 \begin{array}{c}
 0 \\
 -xz \\
 xy \\
 \end{array}
 \right),
\end{eqnarray}
where $x,\,y,\,z$ are the state variables of the system. This
canonical form contains a family of chaotic systems with the same
nonlinear terms, the same symmetry about the $z$-axis, the same
stability of three equilibria, and similar attractors in shape with
a two-scroll and butterfly-like structure

According to \cite{CFCV1994} (see also \cite{CFvanecek1996}), the
Lorenz system satisfies $a_{12}a_{21}>0$ while the Chen system was
found to satisfy $a_{{12}}a_{{21}}<0$. In this sense, the Chen
system is {\it dual\/} to the Lorenz system. In between, it was also
found that there is a transition, the L\"{u} system, which satisfies
$a_{{12}}a_{{21}}=0$ \cite{Alu2002}.

Lately, another form of a unified Lorenz-type system and its
canonical form were developed, which contain some generalized
Lorenz-type systems and their corresponding conjugate Lorenz-type
systems \cite{yang2006,yang2007}.

Moreover, another classification was developed in \cite{yang2008},
where system (3) is classified into two groups, the Lorenz system
group if $a_{11}a_{22}>0$ and the Chen system group if
$a_{11}a_{22}<0$. This classification leads to the finding of a
transition, the Yang-Chen chaotic system, which satisfies
$a_{{11}}a_{{22}}=0$.

Comparing these two classification, one can see that every such
system is classified according to its algebraic structure.
Specifically, it is determined either by $a_{12}a_{21}$ or by
$a_{11}a_{22}$.

Given the above discussions, at this point it is interesting to
ask the following questions:
\begin{enumerate}
 \item Concerning the conditions on the signs of $a_{12}a_{21}$,
defined in \cite{CFCV1994} (see also \cite{CFvanecek1996}), are
these signs essential to the system dynamics? For instance, can a
system belonging to the Lorenz systems family generate Chen-like
attractors?
 \item  Concerning the conditions on the signs of $a_{11}a_{22}$,
defined in \cite{yang2008}, are these signs essential to the system
dynamics? For instance, can a system belonging to the Chen system
group generate Lorenz-like attractors?
 \item Concerning the unified chaotic system (\ref{unified}),
is there any other simpler chaotic system that can also generate
both Lorenz-like attractor and Chen-like attractor?
\end{enumerate}

This paper aims to provide certain answers to the above questions
via the finding of a new family of Lorenz-like systems. This new
family has only one real parameter in a simple algebraic form but
demonstrates very rich and complex dynamics in a way similar to
the generalized Lorenz canonical form
\cite{CFCV2002a,CFCV2002b,CFCV2005}. It belongs to the Lorenz-type
of systems defined in \cite{CFCV1994} and it is also classified
into the Chen system group defined in \cite{yang2008}. However, it
has a very simple form and can generate both visually Lorenz-like
attractors and Chen-like attractors by gradually changing the
single parameter. This reveals that further study of the system
algebraic structure and its effects on the system dynamics remains
an important and interesting challenge.

\section{The new family of chaotic systems}

The new system under investigation is described by
\begin{equation}\label{chenwangeq}
 \left\{
 \begin{array}{l}
 \dot{x}=-x-y \\
 \dot{y}=-x+ry-xz \\
 \dot{z}=-0.1z+xy,
 \end{array}
 \right.
\end{equation}
where $r$ is a real parameter.

This is a one-parameter family of chaotic systems in the sense
that as the real parameter $r$ is gradually varied, a sequence of
chaotic attractors can be continuously generated from the system,
with very rich and complicated dynamics, as further demonstrated
below.

Before proceeding to chaotic dynamics, some basic dynamical
properties of this new system is analyzed following a standard
routine, which is necessary for understanding the nature of the
new system.

\subsection{Symmetry and dissipativity}

First, it is apparent that system (\ref{chenwangeq}) has a natural
symmetry about the $z$-axis under the following transformation:
\[
S(x,y,z)\rightarrow(-x,-y,z).
\]

Second, one may construct the following Lyapunov function:
\begin{equation}
V(x,y,z)=\frac{1}{2}(x^2+y^2+z^2),
\end{equation}
which gives
\begin{eqnarray}
\dot V(x,y,z)&=&x\dot x+y\dot y+z\dot z\nonumber\\
&=&x(-x-y)+y(-x+ry-xz)+z(-0.1z+xy)\nonumber\\
&=&-(x+y)^2+(r+1)y^2-0.1z^2.
\end{eqnarray}
This implies that system (\ref{chenwangeq}) is globally uniformly
asymptotically stable about its zero equilibrium if $r<-1$.
Consequently, system (\ref{chenwangeq}) is not chaotic in the
parameter region $r<-1$.

Third, since
\[
\nabla V=\frac{\partial{\dot x}}{\partial x}+\frac{\partial{\dot
y}}{\partial y} +\frac{\partial{\dot z}}{\partial z}=r-1.1\,,
\]
the system is dissipative under the condition of $r<1.1$. More
precisely, since $V(t)=V(t_0)e^{-(1.1-r)t}$, any initial volume
$V_0$ containing the system trajectories shrinks to zero as $t\to
+\infty$ at an exponential rate of $1.1-r$.

\subsection{Equilibria and their stability}

It is obvious that the origin is a trivial equilibrium of system
(\ref{chenwangeq}). Other nonzero equilibria can be found by
solving the following equations simultaneously:
\[
-x-y=0;\qquad -x+ry-xz=0;\qquad -0.1z+xy=0.
\]

When $r>-1$, system (\ref{chenwangeq}) has three equilibria:
$O(0,0,0)$,
$E_1\left(\sqrt{\frac{1+r}{10}},-\sqrt{\frac{1+r}{10}},-1-r\right)$,
$E_2\left(-\sqrt{\frac{1+r}{10}},\sqrt{\frac{1+r}{10}},-1-r\right)$.

By linearizing system (\ref{chenwangeq}) at $O(0,0,0)$, one obtains the Jacobian
\begin{eqnarray}
J|_O =\left[
  \begin{array}{ccc}
  -1 & -1 & 0 \\
  -1-z & r & -x \\
  y & x & -0.1 \\
  \end{array}
  \right],
\end{eqnarray}
whose characteristic equation is
\[
 {\rm det}(\lambda I-J|_O)
 =\lambda^3+(1.1-r)\lambda^2-(1.1r+0.9)\lambda-0.1(r+1)=0,
\]
which yields
\begin{eqnarray}
\lambda_1&=&-0.1<0,\nonumber\\
\lambda_2&=&-0.5+0.5r+\frac{\sqrt{5+2r+r^2}}{2}>0,\nonumber\\
\lambda_3&=&-0.5+0.5r-\frac{\sqrt{5+2r+r^2}}{2}<0.\nonumber
\end{eqnarray}
Obviously, the equilibrium $O(0,0,0)$ is a saddle point, at which
the stable manifold $W^s(O)$ is two-dimensional and the unstable
manifold $W^u(O)$ is one-dimensional.

Similarly, linearizing the system with respect to the other
equilibria, $E_1$ and $E_2$, yields the following characteristic
equations:
\[
 {\rm det}(\lambda I-J|_{E_{1,2}})
 =\lambda^3+(1.1-r)\lambda^2+0.2\lambda+0.2(r+1)=0.
\]
According to the Routh-Hurwitz criterion, the following
constraints are imposed:
\begin{eqnarray}
 \Delta_1&=&(1.1-r)>0,\\
 \Delta_2&=&\left|
 \begin{array}{cc}
 1.1-r & 0.2(r+1) \\
 1 & 0.2 \\
 \end{array}
 \right|=0.1-2r>0,\\
 \Delta_3&=&0.2(r+1)\Delta_2>0.
\end{eqnarray}
This characteristic polynomial has three roots, all with negative
real parts, under the condition of $-1<r<0.05$. Therefore, the two
equilibria $E_1$ and $E_2$ are both stable nodes, or node-foci, if
$-1<r<0.05$.

However, if the condition $0.05<r<1.1$ holds, then the
characteristic equation has one negative real root and one pair of
complex conjugate roots with a positive real part. Therefore, the
two equilibria $E_{1,2}$ are both saddle-foci, at each of which
the stable manifold $W^s(E_{1,2})$ is one-dimensional and the
unstable manifold $W^u(E_{1,2})$ is two-dimensional.

Summarizing the above analysis and discussions establishes the
following result:

\begin{theorem}
With parameter $-1<r<1.1$, system (\ref{chenwangeq}) has three
equilibria:
\[
 O(0,0,0),
\qquad
 E_1\left(\sqrt{\frac{1+r}{10}},-\sqrt{\frac{1+r}{10}},-1-r\right),
\qquad
 E_2\left(-\sqrt{\frac{1+r}{10}},\sqrt{\frac{1+r}{10}},-1-r\right).
\]
Furthermore,\\
{\rm(i)}\ \ $O(0,0,0)$ is a saddle point, at which the stable
manifold $W^s(O)$ is two-dimensional and the unstable manifold
$W^u(O)$ is one-dimensional;\\
{\rm(ii)}\ if $0.05<r<1.1$, then the two equilibria $E_1$ and
$E_2$ are both saddle-foci, at each of which the stable manifold
$W^s(E_{1,2})$ is one-dimensional and the unstable manifold
$W^u(E_{1,2})$ is two-dimensional;\\
{\rm(iii)} if $-1<r<0.05$, then the equilibria $E_1$ and $E_2$ are
both stable nodes, or node-foci, at each of which the stable
manifold $W^s(E_{1,2})$ is three-dimensional.
\end{theorem}

The Jacobian eigenvalues evaluated at the three equilibria of the
unified chaotic system (\ref{unified}) and of the new system
(\ref{chenwangeq}) are listed in Table 1 for comparison.

\begin{table}[h]
\centering \caption{Equilibria and eigenvalues of several typical
systems.}
{\begin{tabular}{c c c c}\\[-2pt]
\hline
 Systems & Equations & Equilibria & Eigenvalues \\[6pt]
\hline\\[-2pt]
 Unified system & $\dot x=10(y-x)$&$(0,0,0)$ &$-22.8277,-2.6667,11.8277$ \\[1pt]
   $\alpha=0$ &$\dot y=28x-y-xz$&{}&{}\\[2pt]
{} &$\dot z=-\frac{8}{3}z+xy$ &$(\pm6\sqrt{2},\pm6\sqrt{2},27)$
   &$-13.8546,\textbf{0.0940}\pm0.1945i$ \\\hline\\[-2pt]
Unified system&$\dot x=\frac{45}{2}(y-x)$&$(0,0,0)$
   &$-28.1696,-2.8333,19.1696$ \\[1pt]
   $\alpha=0.5$ &$\dot y=\frac{21}{2}x+\frac{27}{2}y-xz$&{}&{}\\[2pt]
{} &$\dot z=-\frac{17}{6}z+xy$ &$(\pm\sqrt{68},\pm\sqrt{68},24)$
   &$-16.9593,\textbf{2.5630}\pm13.1857i$ \\\hline\\[-2pt]
Unified system&$\dot x=30(y-x)$&$(0,0,0)$ &$-30.000,-2.9333,22.200$ \\[1pt]
   $\alpha=0.8$ &$\dot y=\frac{111}{5}y-xz$&{}&{}\\[2pt]
{} & $\dot z=-\frac{44}{15}z+xy$
   &$(\pm\frac{2\sqrt{407}}{5},\pm\frac{2\sqrt{407}}{5},22.2)$
   &$-17.9535,\textbf{3.6101}\pm14.3037i$ \\\hline\\[-2pt]
Unified system&$\dot x=35(y-x)$&$(0,0,0)$ &$-30.8357,-3,23.8359$ \\[1pt]
   $\alpha=1$ &$\dot y=-7x+28y-xz$&{}&{}\\[2pt]
{} &$\dot z=-3z+xy$ &$(\pm3\sqrt{7},\pm3\sqrt{7},21)$
   &$-18.4288,\textbf{4.2140}\pm14.8846i$ \\\hline\hline\\[-2pt]
New system&$\dot x=-x-y$&$(0,0,0)$ &$-0.1,0.6544,-1.6044$ \\[1pt]
   $r=0.05$ &$\dot y=-x+0.05y-xz$&{}&{}\\[2pt]
{} &$\dot z=-0.1z+xy$ &$(\pm\sqrt{42}/20,\mp\sqrt{42}/20,-1.05)$
   &$-1.05,\textbf{0}\pm0.4472i$ \\\hline\\[-2pt]
New system&$\dot x=-x-y$&$(0,0,0)$ &$-0.1,0.6913,-1.5913$ \\[1pt]
   $r=0.1$ &$\dot y=-x+0.1y-xz$&{}&{}\\[2pt]
{} &$\dot z=-0.1z+xy$ &$(\pm\sqrt{11}/10,\mp\sqrt{11}/10,-1.1)$
   &$-1.0162, \textbf{0.0081}\pm0.4652i$ \\\hline\\[-2pt]
New system&$\dot x=-x-y$&$(0,0,0)$ &$-0.1,1,-1.5$ \\[1pt]
   $r=0.5$ &$\dot y=-x+0.5y-xz$&{}&{}\\[2pt]
{} &$\dot z=-0.1z+xy$ &$(\pm\sqrt{15}/10,\mp\sqrt{15}/10,-1.5)$
   &$-0.8101, \textbf{0.1051}\pm0.5994i$ \\\hline\\[-2pt]
New system&$\dot x=-x-y$&$(0,0,0)$ &$-0.1,1.1624,-1.4624$ \\[1pt]
   $r=0.7$ &$\dot y=-x+0.7y-xz$&{}&{}\\[2pt]
{} &$\dot z=-0.1z+xy$ &$(\pm\sqrt{17}/10,\mp\sqrt{17}/10,-1.7)$
   &$-0.7446, \textbf{0.1723}\pm0.6534i$ \\\hline\\[-2pt]
New system&$\dot x=-x-y$&$(0,0,0)$ &$-0.1, 1.2038,-1.4538$ \\[1pt]
   $r=0.75$ &$\dot y=-x+0.75y-xz$&{}&{}\\[2pt]
{} &$\dot z=-0.1z+xy$ &$(\pm\sqrt{70}/20,\mp\sqrt{70}/20,-1.75)$
   &$-0.7312, \textbf{0.1906}\pm0.6651i$ \\[-2pt]
\hline
\end{tabular}}
\end{table}

Given the above analysis, system (\ref{chenwangeq}) will be
discussed only within the parameter region of $0.05<r<1.1$ below.

\subsection{Remarks on the classification of the new system}

On one hand, it is noted that in system (\ref{chenwangeq}) one has
$a_{12}a_{21}=1>0$, so it belongs to the Lorenz system family
defined in \cite{CFCV1994}.

On the other hand, by making the following transformation:
$$
T: (x,y,z)\rightarrow (x,-y,-z),
$$
system (\ref{chenwangeq}) becomes
\begin{equation}\label{transformed}
 \left\{
 \begin{array}{l}
 \dot{x}=-x+y \\
 \dot{y}=x+ry-xz \\
 \dot{z}=-0.1z+xy,
 \end{array}
 \right.
\end{equation}
which satisfies $a_{11}a_{22}<0$ and $a_{11}<0$, thus belong to
the Chen system group according to the classification in
\cite{yang2008} (see Tables 3 and 4 therein).

It is therefore very interesting to see that the new system
(\ref{chenwangeq}) belongs to the same class (either Lorenz-type of
systems defined in \cite{CFCV1994} or Chen system group defined in
\cite{yang2008}), yet can generate both Lorenz-like attractors and
Chen-like attractors by gradually changing the single parameter $r$
from $0.05$ to $0.74$, as further detailed next.

\section{Chaotic behaviors and other complex dynamics}

\begin{figure*}
\centering
\includegraphics[width=6in,height=8.2in]{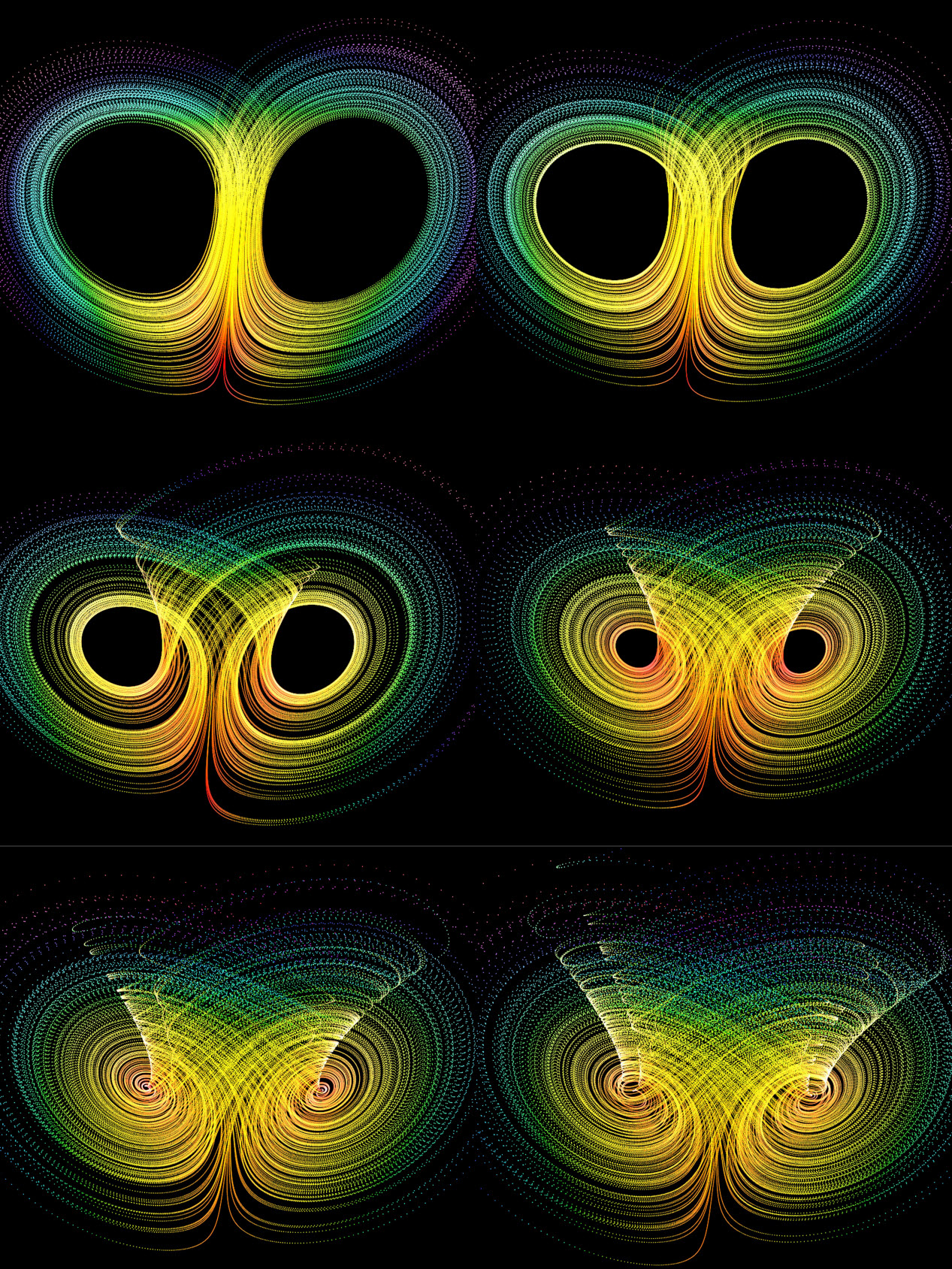}
\caption{Colored figures showing the projected orbit transition
from Lorenz-like to Chen-like attractors in system
(\ref{chenwangeq}).} \label{color}
\end{figure*}

\begin{figure*}
\centering
\includegraphics[width=3.4in,height=2.8in]{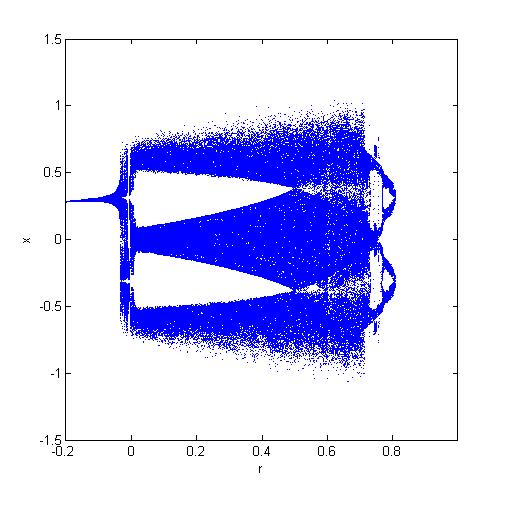}
\caption{Bifurcation diagram of system (\ref{chenwangeq}) with
$r\in[-0.2,0.8]$.} \label{fig:bifur}
\end{figure*}

A complete transition from Lorenz-like to Chen-like attractors in
system (\ref{chenwangeq}) is shown in Fig.\ref{color}. The
bifurcation diagram with respect to parameter $r$ is shown in
Fig.\ref{fig:bifur}. From these figures, one can observe that
system (\ref{chenwangeq}) evolves to a chaotic state and
eventually to a limit cycle, as the parameter $r$ gradually
increases.

To verify the chaotic behaviors of system (\ref{chenwangeq}), its
three Lyapunov exponents $L_1>L_2>L_3$ and the Lyapunov dimension
are calculated, where the latter is defined by
\[
D_L=j+\frac{1}{|L_{j+1}|}\sum_{i=1}^j{L_i},
\]
in which $j$ is the largest integer satisfying
$\sum_{i=1}^j{L_i}\ge0$ and $\sum_{i=1}^{j+1}{L_i}<0$.

\begin{figure*}
\centering
\includegraphics[width=2.9in,height=2.2in]{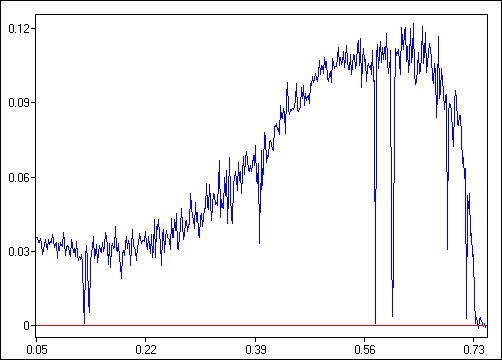}
\caption{The largest Lyapunov exponent of system (\ref{chenwangeq})
with $r\in[0.05,0.75]$.}
\label{fig:newsyslle}
\end{figure*}

Note that system (\ref{chenwangeq}) is chaotic if
$L_1>0,\,L_2=0,\,L_3<0$ with $|L_1|<|L_3|$.
Fig.\ref{fig:newsyslle} shows the dependence of the largest
Lyapunov exponent on the parameter $r$. In particular, for several
values of $r$, the Lyapunov exponents and dimensions of the system
(\ref{chenwangeq}) are summarized in Table 2. From
Fig.\ref{fig:newsyslle}, it is clear that the largest Lyapunov
exponent increases as the parameter $r$ increases from $0.05$ to
$0.7$.

\begin{table}[h]
\centering%
\caption{Numerical results for some values of the parameter $r$ with
initial values $(1,1,1)$.}
{\begin{tabular}{c c c c}\\[-2pt]
\hline
Parameters & Eigenvalues & Lyapunov Exponents & Fractal Dimensions \\[6pt]
\hline\\[-2pt]
$r=0.05$ & $\lambda_1=-1.05$& $L_1=0.03467$ &{}\\[1pt]
{} & $\lambda_{2,3}=\pm0.4472i$ & $L_2=0$ & $D_L=2.0317$ \\[2pt]
{} & {} & $L_3=-1.0843$ & {} \\\hline\\[-2pt]
$r=0.1$ & $\lambda_1=-1.0162$& $L_1=0.03058$ &{}\\[1pt]
{} & $\lambda_{2,3}=0.0081\pm0.4652i$ & $L_2=0$ & $D_L=2.029$ \\[2pt]
{} & {} & $L_3=-1.0301$ & {} \\\hline\\[-2pt]
$r=0.5$ & $\lambda_1=-0.8101$ & $L_1=0.10531$ & {} \\[1pt]
{} & $\lambda_{2,3}=0.1051\pm0.5994i$ & $L_2=0$ & $D_L=2.1497$ \\[2pt]
{} & {} & $L_3=-0.70561$ & {} \\\hline\\[-2pt]
$r=0.7$ & $\lambda_1=-0.7446$ & $L_1=0.08953$ & {} \\[1pt]
{} & $\lambda_{2,3}=-0.1723\pm0.6534i$ & $L_2=0$ & $D_L=2.1832$ \\[2pt]
{} & {} & $L_3=-0.48973$ & {} \\\hline\\[-2pt]
$r=0.75$ & $\lambda_1=-0.7312$ & $L_1=0$ & {} \\[1pt]
{} & $\lambda_{2,3}=0.1906\pm0.6651i$ & $L_2=-0.04626$ & $D_L=1.0063$ \\[2pt]
{} & {} & $L_3=-3.0403$ & {} \\[-2pt]
\hline
\end{tabular}}
\end{table}

\subsection*{Case 1: {\rm $r=0.05$}}

When $r<0.05$, the corresponding system has one saddle and two
stable node-foci (for reference, see \cite{yang2008}). The case of
$r<0.05$ is subtle and somewhat complicated, which will be further
studied elsewhere in the near future.

Here, the discussion starts from $r=0.05$. When $r=0.05$, this new
system has one saddle and two centers. The two centers have the
same set of eigenvalues: one negative real eigenvalue and two
conjugate complex eigenvalues with zero real parts. Nevertheless,
the Lyapunov exponents are $L_1=0.03467$, $L_2=0$, and
$L_3=-1.0843$, and the Lyapunov dimension is $D_L=2.0317$, for
initial values $(1,1,1)$. This convincingly implies that the
system is chaotic.

\subsection*{Case 2: {\rm$r\in(0.05,0.7]$}}

\begin{figure*}
\centering%
\begin{minipage}[b]{0.4\textwidth}
\centering
\includegraphics[width=5cm]{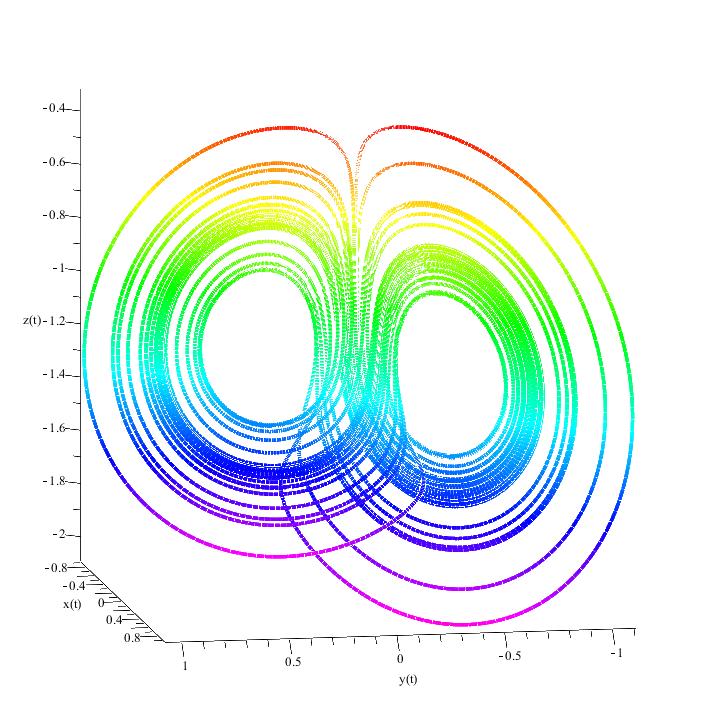}
{\small (a)}
\end{minipage}
\begin{minipage}[b]{0.4\textwidth}
\centering
\includegraphics[width=5cm]{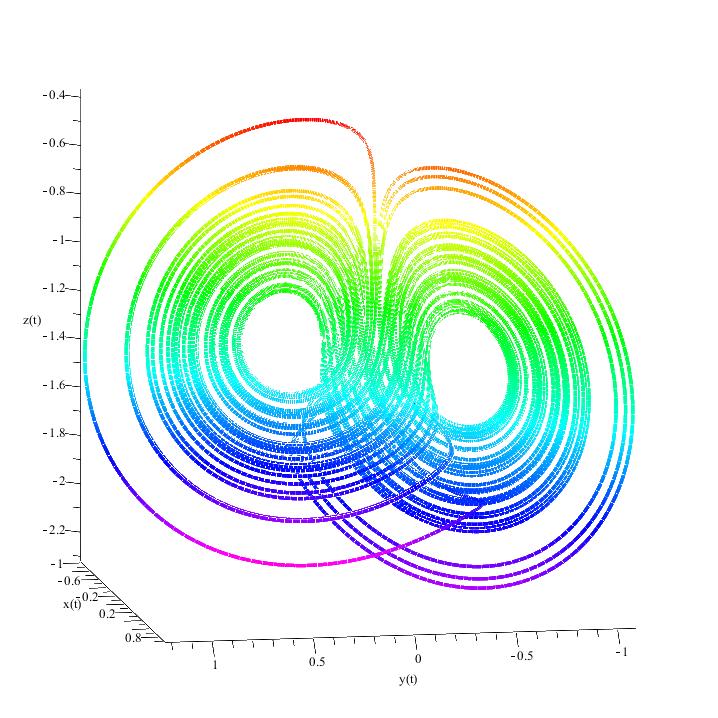}
{\small (b)}
\end{minipage}
\begin{minipage}[b]{0.4\textwidth}
\centering
\includegraphics[width=5cm]{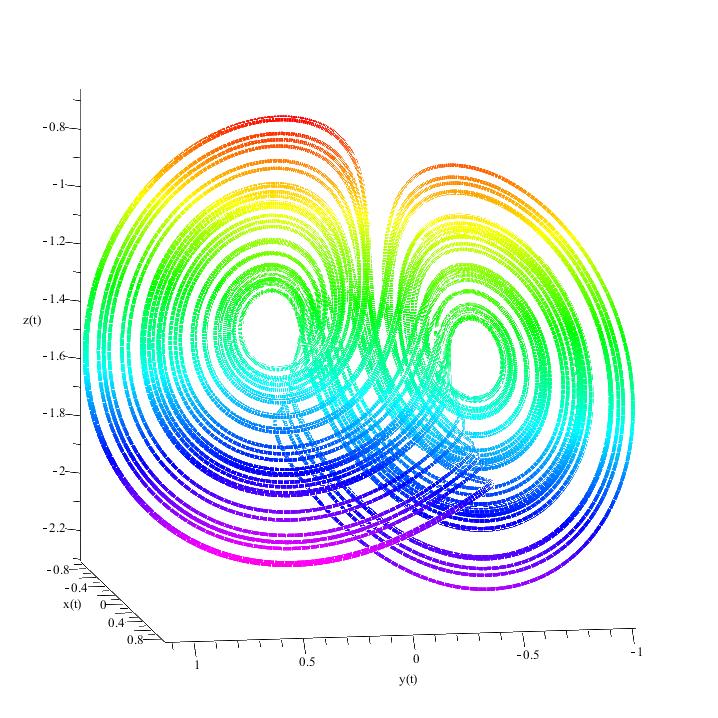}
{\small (c)}
\end{minipage}
\begin{minipage}[b]{0.4\textwidth}
\centering
\includegraphics[width=5cm]{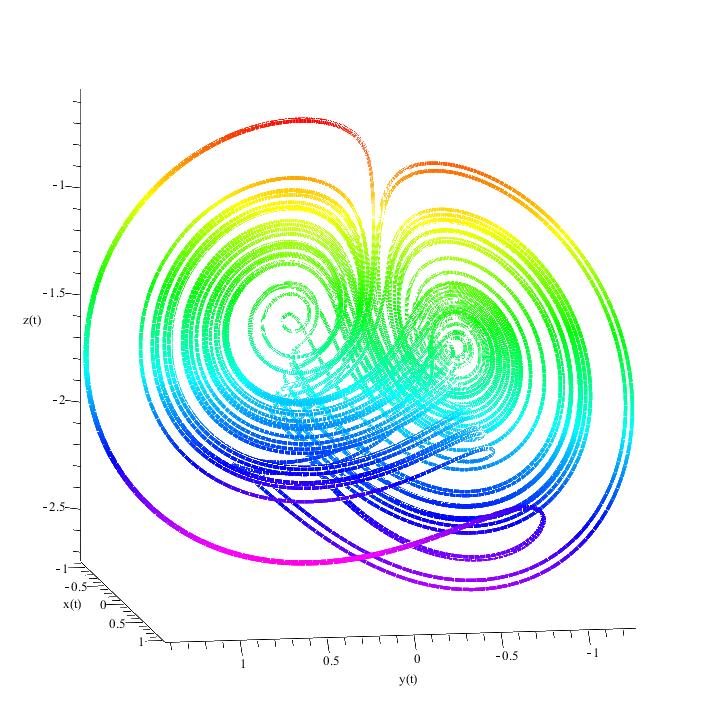}
{\small (d)}
\end{minipage}
\begin{minipage}[b]{0.4\textwidth}
\centering
\includegraphics[width=5cm]{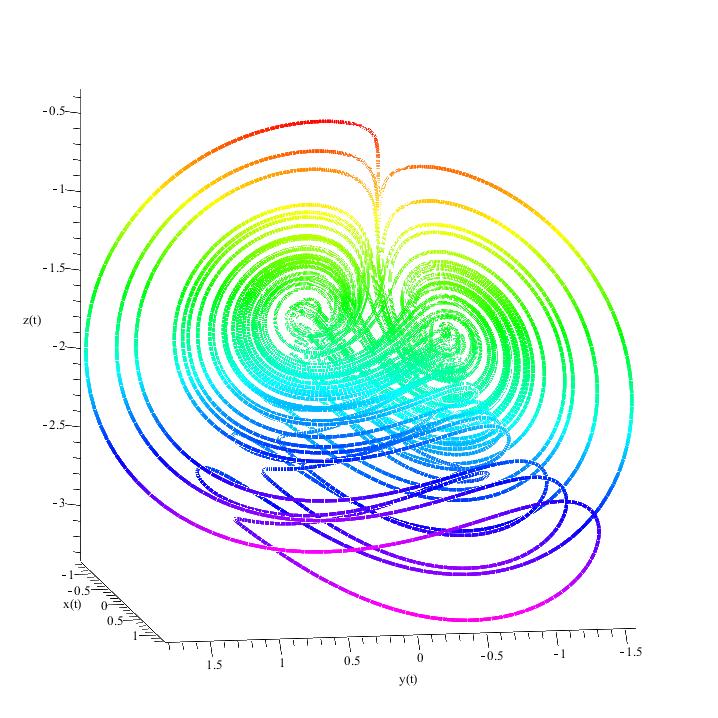}
{\small (e)}
\end{minipage}
\begin{minipage}[b]{0.4\textwidth}
\centering
\includegraphics[width=5cm]{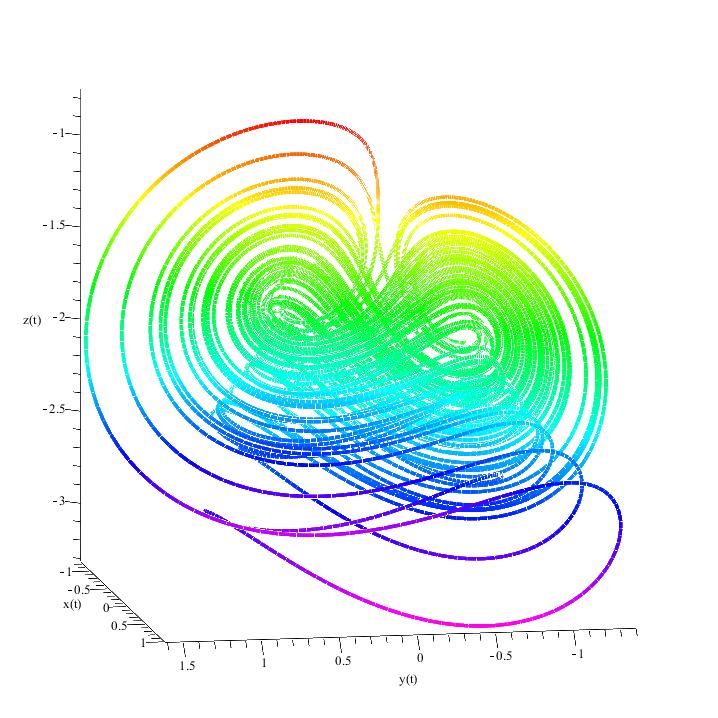}
{\small (f)}
\end{minipage}
\begin{minipage}[b]{0.4\textwidth}
\centering
\includegraphics[width=5cm]{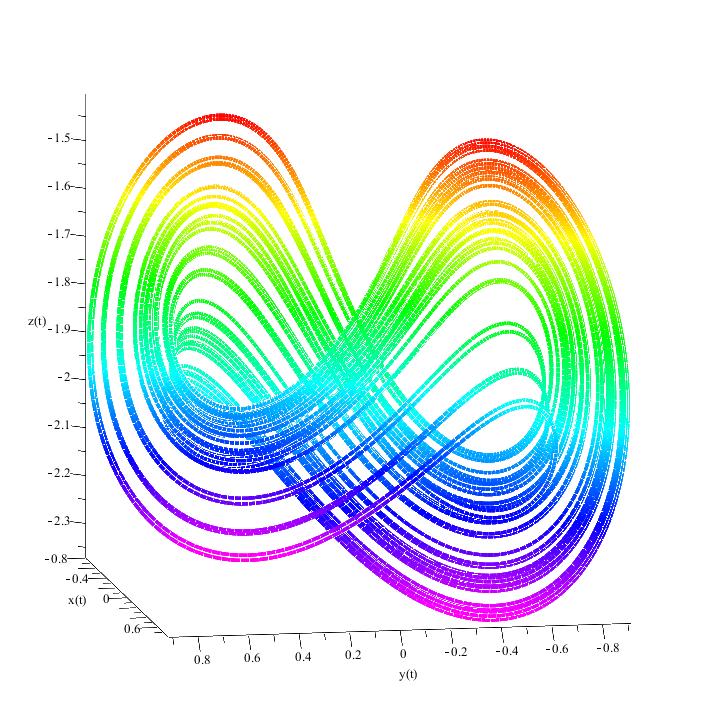}
{\small (g)}
\end{minipage}
\begin{minipage}[b]{0.4\textwidth}
\centering
\includegraphics[width=5cm]{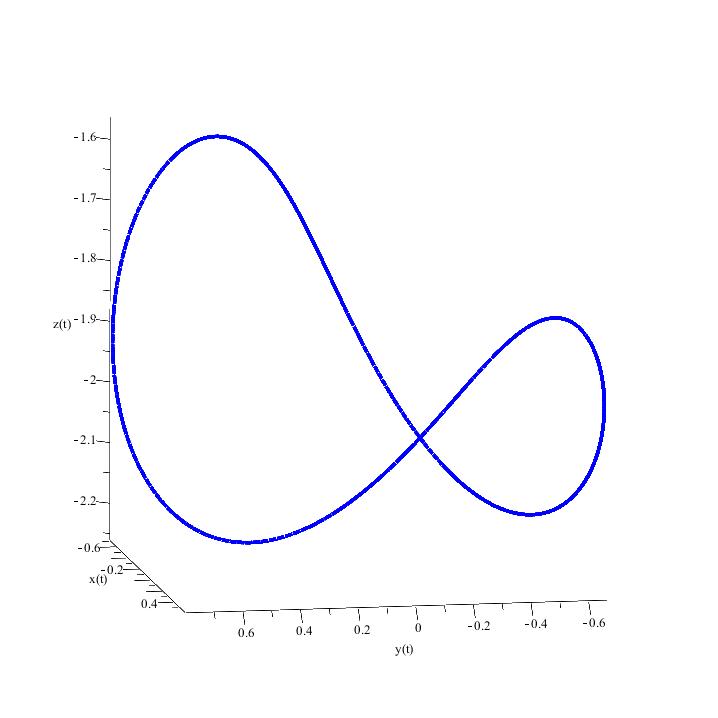}
{\small (h)}
\end{minipage}
\caption{Attractors of system (\ref{chenwangeq}) with parameter values:
(a) $r=0.2$; (b) $r=0.3$; (c) $r=0.4$; (d) $r=0.50$; (e) $r=0.6$;
(f) $r=0.7$; (g) $r=0.72$; (h) $r=0.75$.}
\label{fig:newsys}
\end{figure*}

In this case, the new system has one saddle and two saddle-foci.
The two saddle-foci have one negative real eigenvalue and two
conjugate complex eigenvalues with a positive real part. Moreover,
the largest Lyapunov exponent is larger than zero, which is
increased as the parameter $r$ is increased from $0.05$ to $0.7$.
At the same time, the attractor of this system is changed from
Lorenz-like to Chen-like as the parameter $r$ is increased from
$0.05$ to around $0.7$, as seen in Fig.\ref{fig:newsys}(a)-5(f).

\subsection*{Case 3: {\rm$r\in(0.7,0.74)$}}

When $r$ lies in this interval, system (\ref{chenwangeq}) is still
chaotic, for Fig.\ref{fig:newsyslle} shows that the largest
Lyapunov exponent is positive. However, the attractor is neither
Lorenz-like nor Chen-like, as shown in Fig.\ref{fig:newsys}(g),
when $r=0.72$. Moreover, the chaotic property becomes less
significant as compared with Case 2. Indeed, the largest Lyapunov
exponent decreases when the parameter is increased from $r=0.7$ to
$r=0.74$, as shown in Fig.\ref{fig:newsyslle}.

\subsection*{Case 4: {\rm$r\in[0.74,1]$}}

In this case, although the new system has one saddle and two
saddle-foci, associated with one negative real eigenvalue and two
conjugate complex eigenvalues with a positive real part, the
system is not chaotic. Instead, the system has a limit cycle,
though different from Case 2 and Case 3. Fig.\ref{fig:newsys}(h)
displays the limit cycle when $r=0.75$.

\section{Comparison between the new family and the unified system}

\begin{figure*}
\centering%
\begin{minipage}[b]{0.4\textwidth}
\centering
\includegraphics[width=5cm]{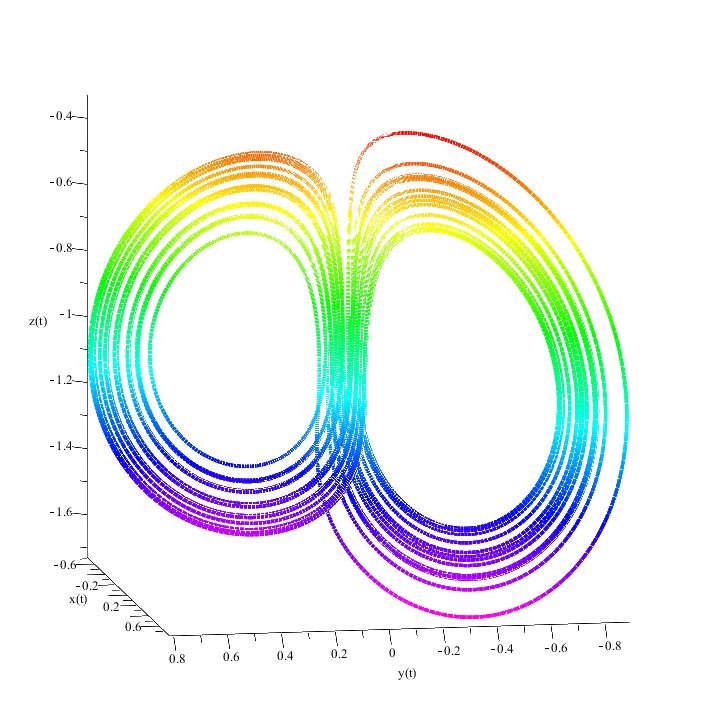}
{\small (a)}
\end{minipage}
\begin{minipage}[b]{0.4\textwidth}
\centering
\includegraphics[width=5cm]{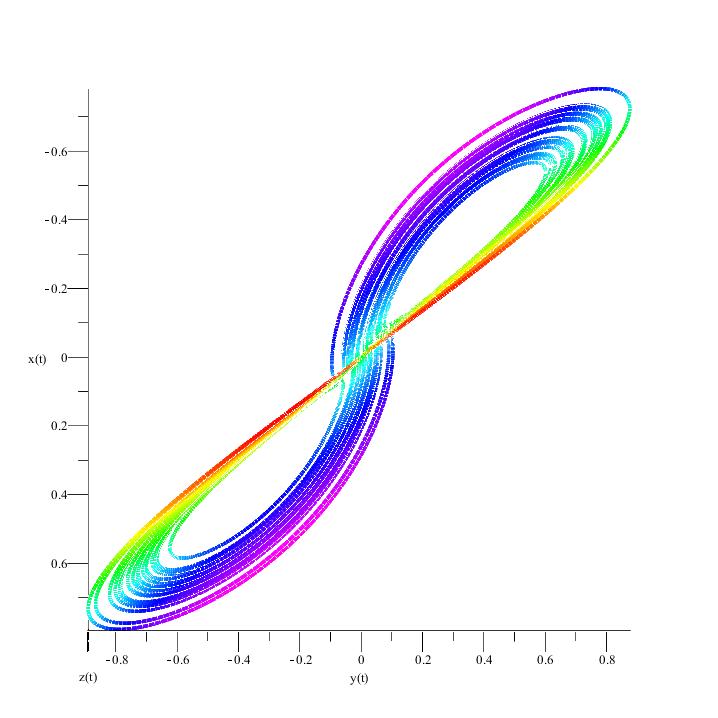}
{\small (b)}
\end{minipage}
\begin{minipage}[b]{0.4\textwidth}
\centering
\includegraphics[width=5cm]{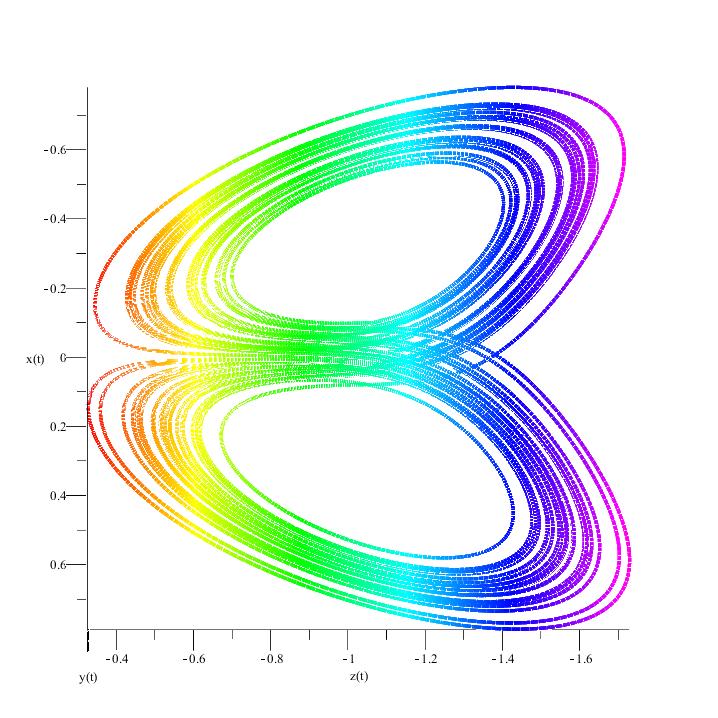}
{\small (c)}
\end{minipage}
\begin{minipage}[b]{0.4\textwidth}
\centering
\includegraphics[width=5cm]{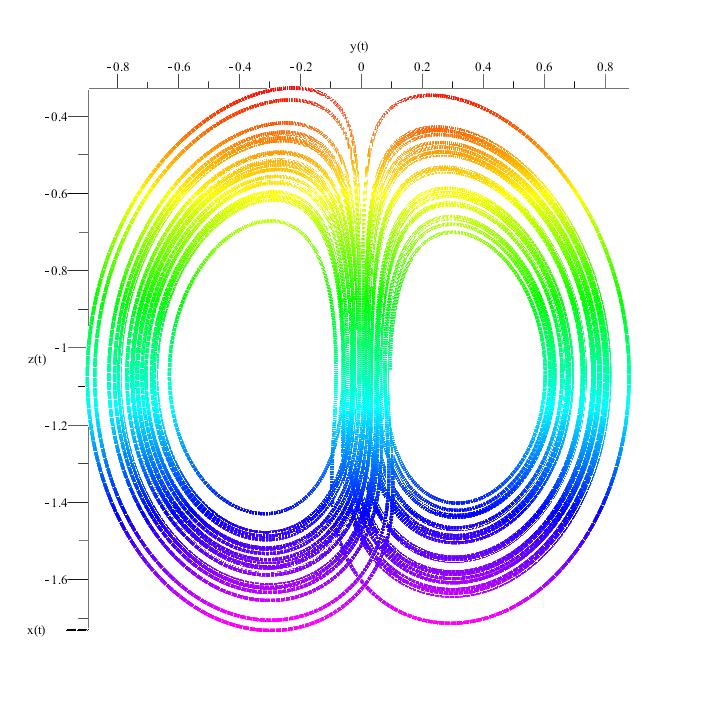}
{\small (d)}
\end{minipage}
\begin{minipage}[b]{0.4\textwidth}
\centering
\includegraphics[width=5cm]{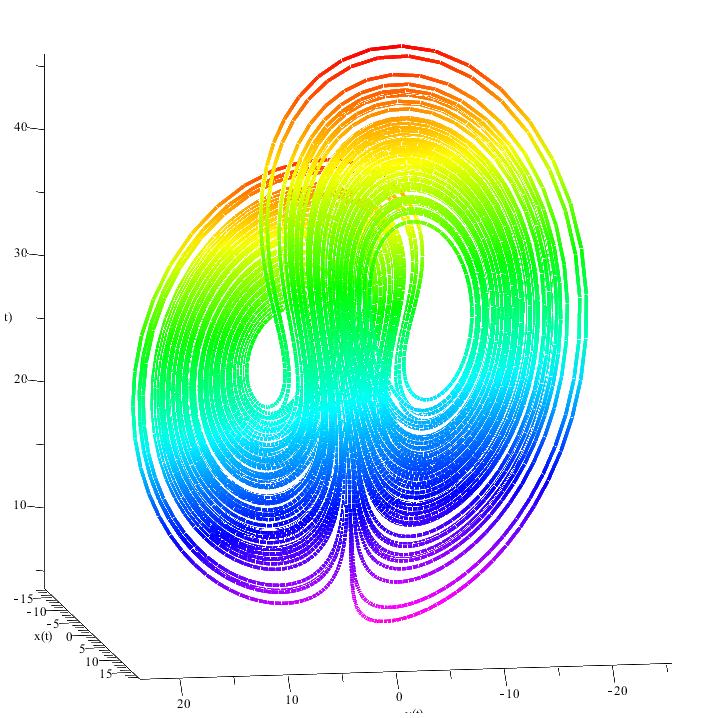}
{\small (e)}
\end{minipage}
\begin{minipage}[b]{0.4\textwidth}
\centering
\includegraphics[width=5cm]{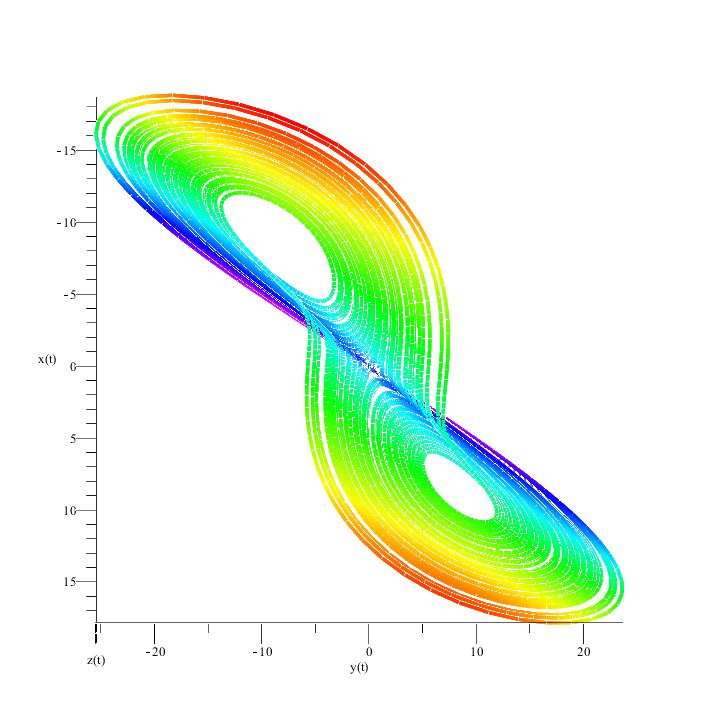}
{\small (f)}
\end{minipage}
\begin{minipage}[b]{0.4\textwidth}
\centering
\includegraphics[width=5cm]{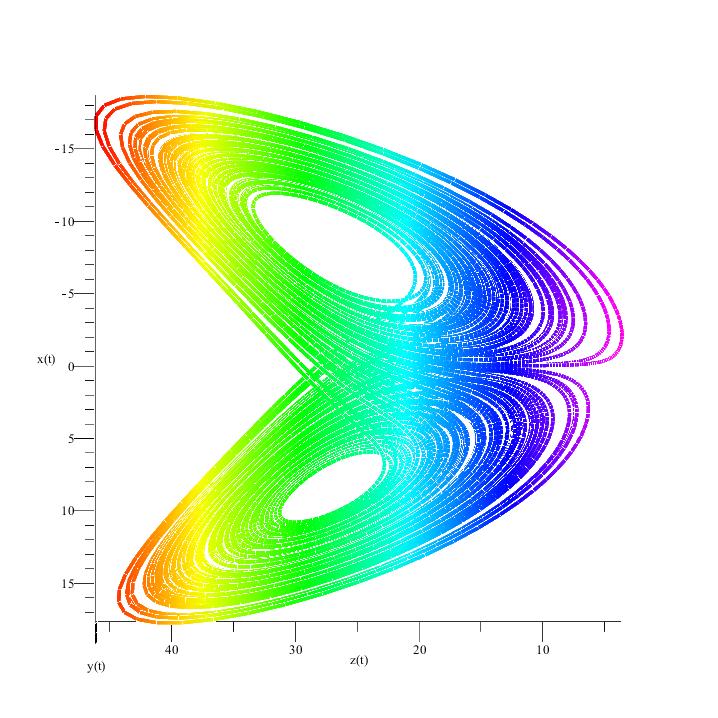}
{\small (g)}
\end{minipage}
\begin{minipage}[b]{0.4\textwidth}
\centering
\includegraphics[width=5cm]{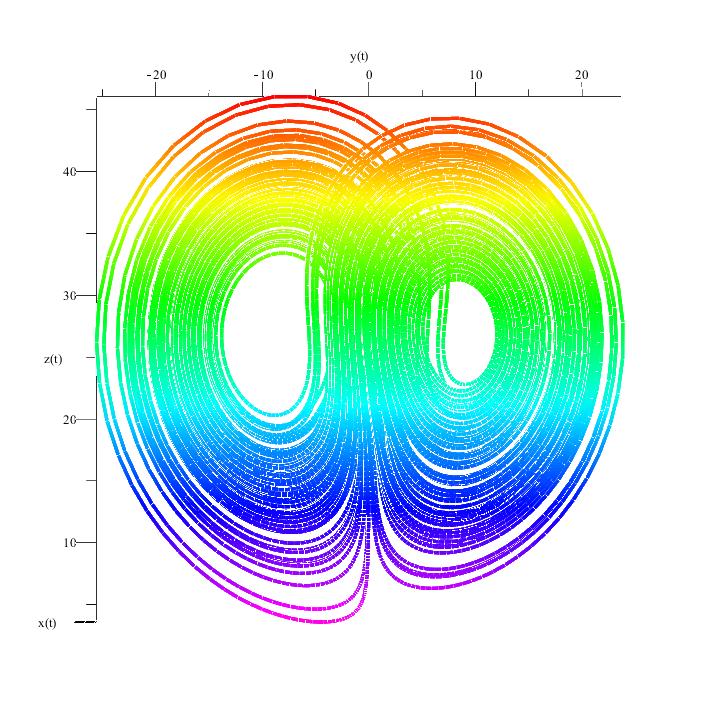}
{\small (h)}
\end{minipage}
\caption{Attractor of system (\ref{chenwangeq}) with $r=0.06$, (a)
the 3D phase portrait; (b) projected orbit on $x$-$y$ plane; (c)
projected orbit on $x$-$z$ plane; (d) projected orbit on $y$-$z$
plane. The Lorenz attractor corresponding to $\alpha=0$ in the
system (\ref{unified}), (e) the 3D phase portrait; (f) projected
orbit on $x$-$y$ plane; (g) projected orbit on $x$-$z$ plane; (h)
projected orbit on $y$-$z$ plane.} \label{fig:lorenz-old-new}
\end{figure*}

\begin{figure*}
\centering%
\begin{minipage}[b]{0.4\textwidth}
\centering
\includegraphics[width=5cm]{SLF3D_70}
{\small (a)}
\end{minipage}
\begin{minipage}[b]{0.4\textwidth}
\centering
\includegraphics[width=5cm]{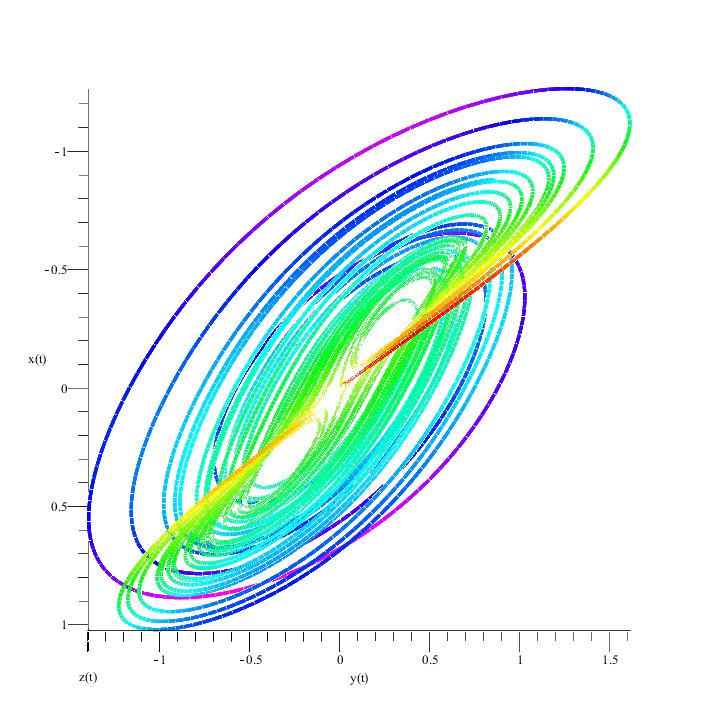}
{\small (b)}
\end{minipage}
\begin{minipage}[b]{0.4\textwidth}
\centering
\includegraphics[width=5cm]{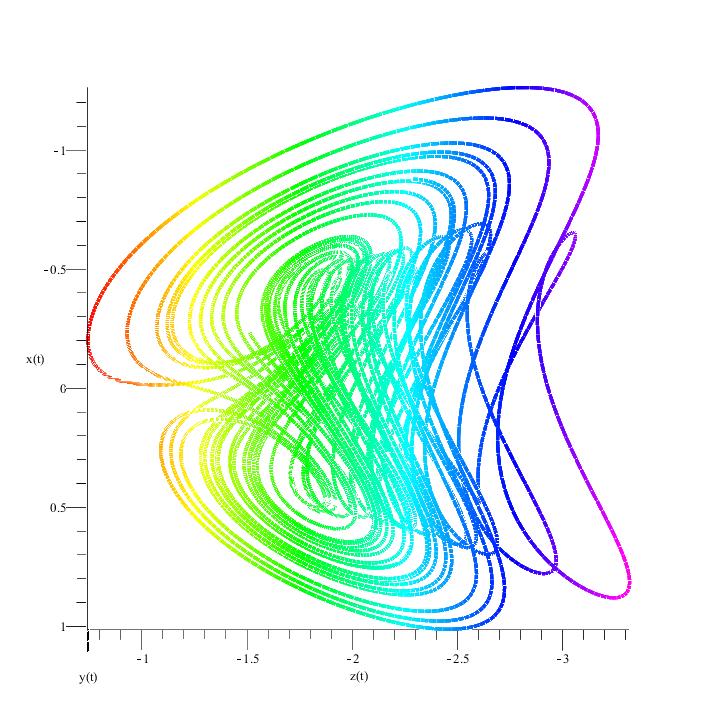}
{\small (c)}
\end{minipage}
\begin{minipage}[b]{0.4\textwidth}
\centering
\includegraphics[width=5cm]{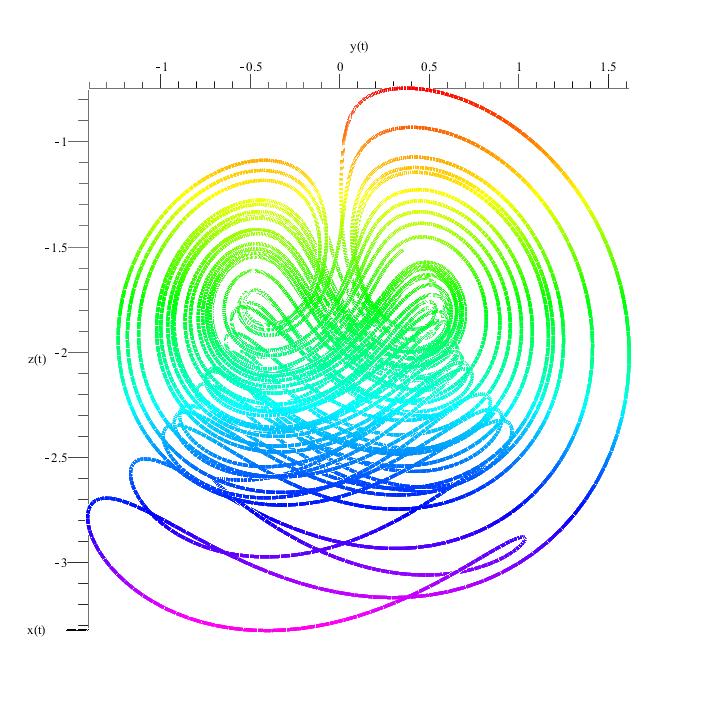}
{\small (d)}
\end{minipage}
\begin{minipage}[b]{0.4\textwidth}
\centering
\includegraphics[width=5cm]{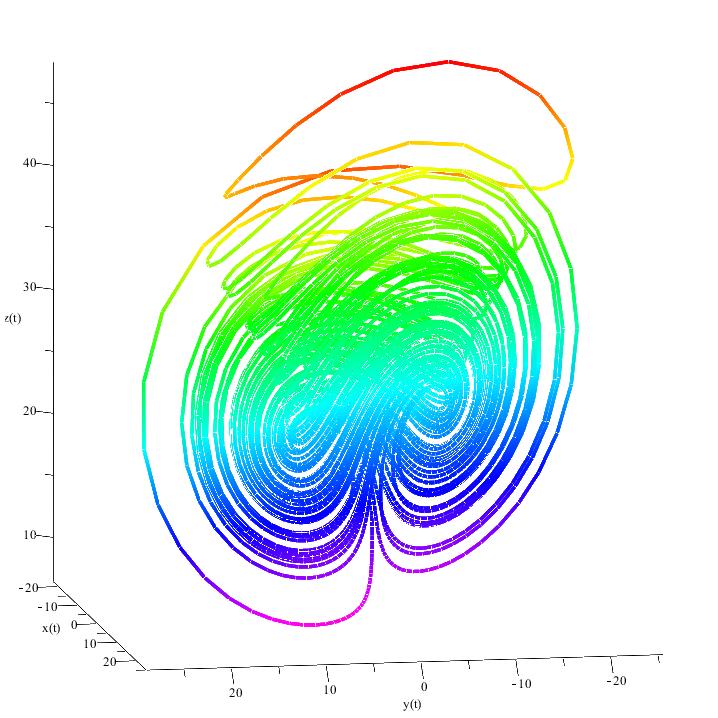}
{\small (e)}
\end{minipage}
\begin{minipage}[b]{0.4\textwidth}
\centering
\includegraphics[width=5cm]{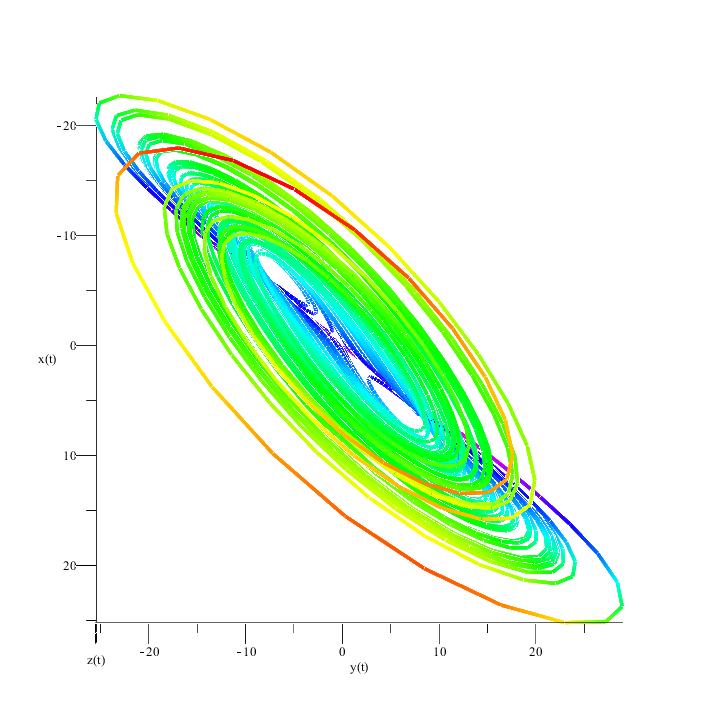}
{\small (f)}
\end{minipage}
\begin{minipage}[b]{0.4\textwidth}
\centering
\includegraphics[width=5cm]{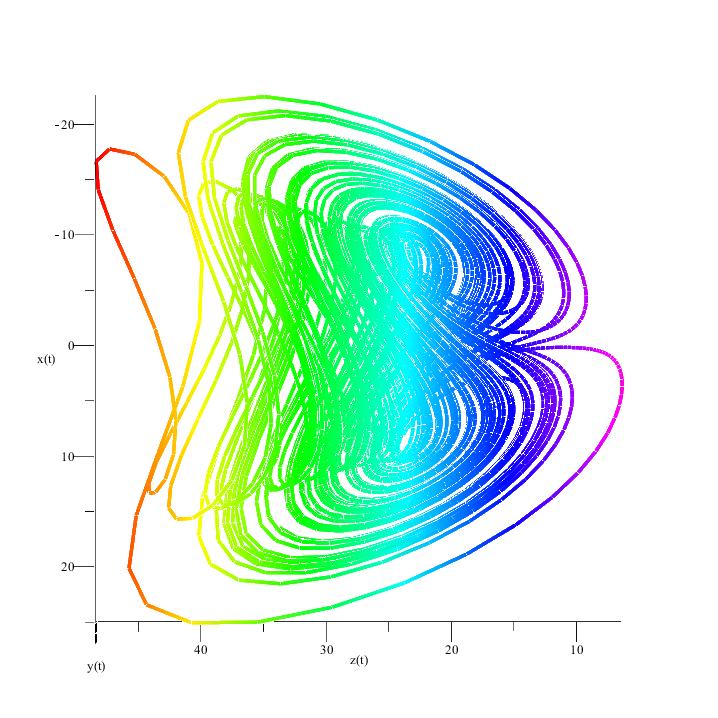}
{\small (g)}
\end{minipage}
\begin{minipage}[b]{0.4\textwidth}
\centering
\includegraphics[width=5cm]{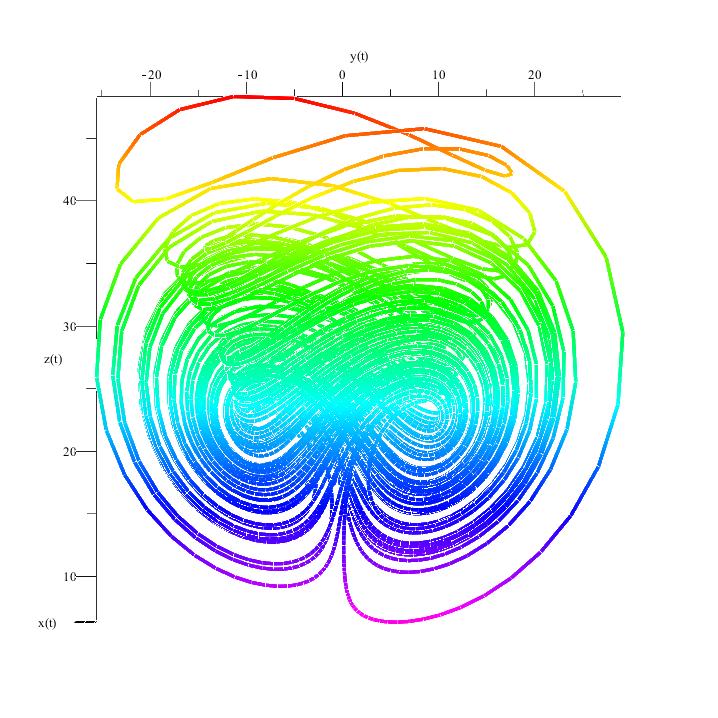}
{\small (h)}
\end{minipage}
\caption{Attractor of system (\ref{chenwangeq}) with $r=0.7$, (a)
the 3D phase portrait; (b) projected orbit on $x$-$y$ plane; (c)
projected orbit on $x$-$z$ plane; (d) projected orbit on $y$-$z$
plane. The Chen attractor corresponding to $\alpha=1$ in the
system (\ref{unified}), (e) the 3D phase portrait; (f) projected
orbit on $x$-$y$ plane; (g) projected orbit on $x$-$z$ plane; (h)
projected orbit on $y$-$z$ plane.} \label{fig:chen-old-new}
\end{figure*}

It is interesting to compare the new family of chaotic systems
with the now-familiar generalized Lorenz systems family,
particularly the corresponding Lorenz-type and Chen-type systems,
as shown in Figs.\ref{fig:lorenz-old-new} and
\ref{fig:chen-old-new}.

\begin{figure*}
\centering
\includegraphics[width=0.45\textwidth]{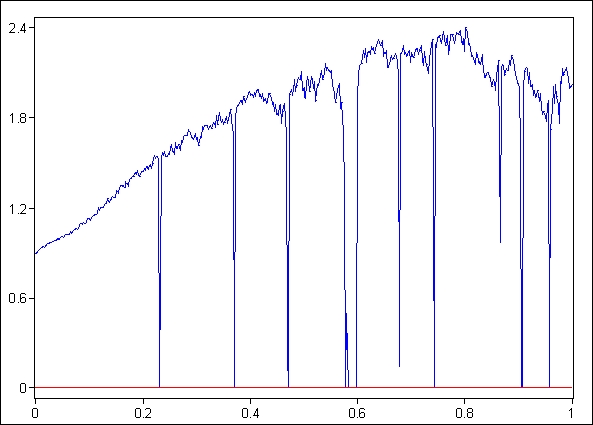}
\caption{The largest Lyapunov exponent of the unified chaotic system
with $\alpha\in[0,1]$.}
\label{fig:unifiedlle}
\end{figure*}

Recall in particular the unified chaotic system (\ref{unified}),
which is chaotic for all $\alpha\in[0,1]$, with a positive largest
Lyapunov exponent as shown in Fig.\ref{fig:unifiedlle}.

There are several common features between the new family of
chaotic systems (\ref{chenwangeq}) and the unified chaotic system
(\ref{unified}):
\begin{enumerate}
  \item They are both one-parameter family of chaotic systems,
  representing infinitely many non-equivalent chaotic systems.
  \item They both demonstrate a continuously changing process
  generating from Lorenz-like to Chen-like attractors.
  \item They both have a similar simple algebraic structure with
  two same nonlinear terms (i.e., $-xz$ in the second equation and
  $xy$ in the third equation), juts like the Lorenz and the Chen
  systems.
  \item They both have the same $z$-axis rotational symmetry, therefore
  all their attractors are visually similar in shape with a two-scroll
  and butterfly-like structure.
  \item They both have three equilibria: one saddle and two saddle-foci.
  Moreover, the two saddle-foci have the same eigenvalues: one negative
  real number and two conjugate complex numbers with a positive real part.
  Specifically, for the unified chaotic system, the positive real part
  of the conjugate complex eigenvalues increases from the Lorenz system
  $(0.0940)$ to the Chen system $(4.2140)$, while for the new system,
  it increases as $r$ is increased from $0.05$ to $0.7$.
\end{enumerate}

The main difference between these two one-parameter families of
chaotic systems are quite subtle. For system (\ref{chenwangeq}),
the real part of its conjugate complex eigenvalues is precisely
zero when $r=0.05$. This is a critical point when the stability of
the equilibrium pairs is changed from stable to unstable.
Therefore, one may define the value of $r=0.05$ as a boundary, or
more precisely as a starting point of the Lorenz systems family.
Since the parameter $r$ can take values on both sides of $r=0.05$
to generate chaos, the new family system (\ref{chenwangeq}) is
richer and more interesting than the unified system system
(\ref{unified}).

\section{Concluding remarks and future research}

\subsection{Concluding remarks}

Regarding the few questions raised in the Introduction section, we
now come out with some answers:
\begin{enumerate}
 \item In this paper, we have reported the finding of a structurally
simple yet dynamically complex one-parameter family of 3D
autonomous systems with only two quadratic terms like the Lorenz
system, with some key constant coefficients set as $1$ or $-1$,
leaving only one tunable real parameter in the linear part of the
system.
 \item By tuning the singe parameter, this new family of systems can
generate a chain of chaotic attractors in a very similar gradual
changing process like the one from Lorenz attractor to Chen
attractor generated by the generalized Lorenz systems family and
the one by the unified chaotic system.
 \item Unlike the typical Chen system, however, the new system that
generates the Chen-like attractor does not satisfy the condition
$a_{12}a_{21} < 0$ in the linear part of the generalized Lorenz
canonical form defined in \cite{CFCV1994}.
 \item Unlike the classical Lorenz system, the new system that can
generates a similar Lorenz-like attractor does not satisfy the
condition $a_{11}a_{22}>0$ in the linear part of the generalzied
Lorenz canonical form defined in \cite{yang2008}.
 \item This system is richer and more interesting than the unified
chaotic system, in the sense that the new single parameter has a
wider range to generate similar form of chaos.
 \item Similar to the generalized Lorenz systems family, the new
system also has three equilibria, with one saddle and two
saddle-foci, where the two saddle-foci have the same eigenvalues
(one is negative real and two are complex conjugate with a
positive real part). However, the in the new system the real part
of the eigenvalues starts from zero, which can be used to define a
starting point for the Lorenz-like systems family. In this sense,
the new system has literally extended the unified chaotic system,
even the generalized Lorenz systems family, to some extent.
\end{enumerate}

\subsection{Future research}

Some related research issues may be further pursued in the near
future:
\begin{enumerate}
 \item There may exist other Lorenz-type systems with the same
nonlinear terms as the classical Lorenz system, which can generate
a variety of chaotic attractors.
 \item There may even exist other types of nonlinear terms that
satisfy the expected $z$-axis rotational symmetry, for which the
nonlinear term in the second equation must be like $xz$ or $yz$,
and that in the third equation must be like $xy$, or $z^{2}$, or
$x^{2}$, or $y^{2}$. So, if one wants to study 3D autonomous
systems with two quadratic terms that can maintain the $z$-axis
rotational symmetry, then there are totally $2\times4=8$ possible
combinations to consider. It is therefore interesting to ask if
there would be other chaotic systems with two nonlinear terms
different from those studied so far around the Lorenz system.
 \item In the new family of chaotic systems studied in this paper,
the real part of the complex conjugate eigenvalues at system
equilibria starts from zero for the case with $r=0.05$, which can
be defined as the starting point of the Lorenz-like systems
family. This particularly interesting critical system with
$r=0.05$ alone deserves further investigations.
 \item This paper has further revealed that some systems with
different structures can literally generate similar-shaped chaotic
attractors. Therefore, for 3D autonomous systems with two
quadratic terms, the relation between the system algebraic
structure and system chaotic dynamics is an important and
interesting issue to be further revealed, understood and analyzed.
\end{enumerate}

\section*{Acknowledgements}
The authors sincerely thank Professors Sergej \v{C}elikovsk\'{y},
Qigui Yang and Tianshou Zhou for their valuable comments and
suggestions. This work was supported by the France/HongKong Joint
Research Scheme under grant CityU 9052002/2011-12.


\begin{thebibliography}{10}

\bibitem{Alorenz}
Edward~N. Lorenz.
\newblock Deterministic nonperiodic flow.
\newblock {\em Journal of the Atmospheric Sciences}, 20(2):130--141, 1963.

\bibitem{Alorenz2}
C.~Sparrow.
\newblock {\em The Lorenz equations: Bifurcation, Chaos, and Strange
  Attractor}.
\newblock Springer-Verlag, New York, 1982.

\bibitem{Achen1999}
G.~Chen and T.~Ueta.
\newblock Yet another chaotic attractor.
\newblock {\em Int. J. Bifur. Chaos}, 9:1465--1466, 1999.

\bibitem{Aueta2000}
T.~Ueta and G.~Chen.
\newblock Bifurcation analysis of chen's equation.
\newblock {\em Int. J. Bifur. Chaos}, 10:1917--1931, 2000.

\bibitem{AChenexists}
T.~Zhou, Y.~Tang, and G.~Chen.
\newblock Chen's attractor exists.
\newblock {\em Int. J. Bifur. Chaos}, 14:3167--3177, 2004.

\bibitem{Achendynamical2003}
T.~Zhou, Y.~Tang, and G.~Chen.
\newblock Complex dynamical behaviors of the chaotic chen's system.
\newblock {\em Int. J. Bifur. Chaos}, 13:2561--2574, 2003.

\bibitem{Alu2002b}
J.~L\"{u}, G.~Chen, D.~Cheng, and S.~Celikovsky.
\newblock {Bridge the gap between the Lorenz system and the Chen system}.
\newblock {\em Int. J. Bifur. Chaos}, 12(12):2917--2926, 2002.

\bibitem{CFCV1994}
S.~\v{C}elikovsk\'{y} and A.~Vane\v{c}\v{e}k.
\newblock Bilinear systems and chaos.
\newblock {\em Kybernetika}, 30:403--424, 1994.

\bibitem{CFCV2002a}
S.~\v{C}elikovsk\'{y} and G.~Chen.
\newblock On a generalized lorenz canonical form of chaotic systems.
\newblock {\em Int. J. Bifur. Chaos}, 12:1789--1812, 2002.

\bibitem{CFCV2002b}
S.~\v{C}elikovsk\'{y} and G.~Chen.
\newblock Hyperbolic-type generalized lorenz system and its canonical form.
\newblock In {\em Proc. 15th Triennial World Congress of IFAC}, Barcelona,
  Spain.

\bibitem{CFCV2005}
S.~\v{C}elikovsk\'{y} and G.~Chen.
\newblock On the generalized lorenz canonical form.
\newblock {\em Chaos Solit. Fract.}, 26:1271--1276, 2005.

\bibitem{CFvanecek1996}
A.~Vane\v{c}\v{e}k and S.~\v{C}elikovsk\'{y}.
\newblock {\em Control systems: From linear analysis to synthesis of chaos}.
\newblock Prentice-Hall, London, 1996.

\bibitem{Alu2002}
J.~L\"{u} and G.~Chen.
\newblock {A new chaotic attractor coined}.
\newblock {\em Int. J. Bifur. Chaos}, 12(3):659--661, 2002.

\bibitem{yang2006}
Q.~Yang, G.~Chen, and T.~Zhou.
\newblock {A unified Lorenz-type system and its canonical form}.
\newblock {\em Int. J. Bifur. Chaos}, 16(10):2855, 2006.

\bibitem{yang2007}
Q.~Yang, G.~Chen, and K.~Huang.
\newblock {Chaotic attractors of the conjugate Lorenz-type system}.
\newblock {\em Int. J. Bifur. Chaos}, 17(11):3929--3949, 2007.

\bibitem{yang2008}
Q.~Yang and G.~Chen.
\newblock {A chaotic system with one saddle and two stable node-foci}.
\newblock {\em Int. J. Bifur. Chaos}, 18:1393--1414, 2008.

\end{thebibliography}

\end{document}